\documentclass[twocolumn,prl]{revtex4}

\usepackage{graphicx}
\usepackage{dcolumn}
\usepackage{bm}

\begin{document}
\title{Possible vortex splitting in the cuprate superconductors}

\author{R. Hlubina}

\affiliation{
Division of Solid State Physics, Comenius University,
Mlynsk\'{a} Dolina F2, 842 48 Bratislava, Slovakia\\
Centre of Excellence of the Slovak Academy of Sciences CENG}

\begin{abstract}
We propose that the observed splitting of the vortices in the cuprates
into fractional vortices (partons) may be of static rather than of
dynamic origin. This interpretation is backed by a study of a model
with a dominant $d$-wave and subdominant $s$-wave pairing
interaction. We find that the vortex may split into two partons, both
of which carry one half of the magnetic flux quantum. The partons are
hold together by a confining string along which the phase jumps
approximately by $\pi$ and their equilibrium distance increases with
lowering the energy difference $\varepsilon$ between the pairing
states. The partons become deconfined at the critical point where
$\varepsilon$ vanishes.
\end{abstract}
\pacs{73.43.Nq, 74.20.Rp, 74.25.-q, 74.72.-h, 74.81.Fa}
\maketitle

The nonsuperconducting phase of the high temperature superconductors
exhibits anomalous features \cite{Lee06}. On the other hand, the
low-temperature superconducting state is believed to be well described
by the standard Bardeen-Cooper-Schrieffer paradigm, if the $d$-wave
pairing symmetry and the Landau Fermi liquid corrections are taken
into account. In view of the anomalies of the normal phase, it is
tempting to look for unconventional features in the superconducting
phase as well. Interesting results have in fact been obtained by the
scanning tunneling spectroscopy of the vortices in
Bi$_2$Sr$_2$CaCu$_2$O$_{8+\delta}$ \cite{Hoogenboom00,Levy05},
according to which the vortex cores may split into several
subcomponents with a spacing in the range 10 - 100~\AA. The
experimental results were interpreted as a result of the vortex
hopping between different pinning sites \cite{Hoogenboom00,Fischer07}.
The purpose of this paper is to propose an alternative interpretation
of the experimental results, in which the splitting of the vortices
into partons is considered to be of static origin.  Based on an
analogy with the physics of dislocations, we will show that the vortex
may split into two fractional vortices (partons), each of them
carrying one half of the magnetic flux quantum $\Phi_0$.

It is well known that screw dislocations in fcc materials can split
into two Shockley partial dislocations whose (fractional) Burgers
vectors add up to an integer lattice vector \cite{Hirth82}.  In that
case the singular dislocation line transforms into a singular strip
whose borders are formed by the partial dislocations. The relative
displacement of the crystal on both sides of the strip is not equal to
a lattice vector.  The dynamical reasons for the stability of the
partial dislocations are: (i) repulsion between the parallel partial
dislocations and (ii) the low elastic energy cost of the displacement
across the strip. In the vortex case we will show that in addition to
(i), which is always true, the criterion (ii) may be satisfied in
superconductors with sufficiently strong subleading pairing
interactions.

Besides serving as an alternative explanation of the experiments
\cite{Hoogenboom00,Fischer07}, the parton hypothesis provides
additional support to the interpretations of the pseudogap in the
high-temperature superconductors as an incoherent liquid of singlet
electron pairs on the bonds of the CuO$_2$ lattice
\cite{Anderson87}. The major open problem in this line of thinking is
the question about the mechanism leading to the phase disordering of
the pairs. It has been argued that in order to destroy the phase
ordering and to stabilize the pseudogap state, the presence of cheap
vortices in the cuprates is required \cite{Lee06}.  We will show that
the energy of the split vortices may be substantially reduced with
respect to the conventional vortex energy.

Our main assumption is the existence, in addition to the leading
$d$-wave interactions, of subleading pairing interactions in the
$s$-wave sector. Our motivation is as follows. It seems reasonable to
assume that the model of the cuprates should contain a strong on-site
repulsion and a moderate antiferromagnetic nearest neighbor spin-spin
interaction. It is well known that within this type of a model,
condensates with both $d$-wave and $s$-wave symmetry may form
\cite{Kotliar88}.  On the other hand, we are not aware of direct
experimental evidence for such subleading pairing tendencies in the
cuprates.  It has been argued, however, that the large second
harmonics of the current-phase relation observed in the cuprate grain
boundary Josephson junctions \cite{Ilichev99,Ilichev01} provides an
indirect evidence for the existence of subleading pairing interactions
\cite{Hlubina03}.

Previous works have noted that the structure of isolated vortex lines
in superconductors with competing pairing interactions may become very
rich. In particular, within phenomenological Ginzburg-Landau theory it
has been shown that in the vicinity of the vortex cores in $d$-wave
superconductors, there may nucleate a finite $s$-wave component with a
nontrivial phase structure \cite{Franz96}.  However, these results do
not explain the experiment \cite{Hoogenboom00}, since the dominant
$d$-wave field has only one singularity. For the same reason, neither
the more recently proposed nonsingular vortices \cite{Melnikov00} can
explain the experimental data. Also Volovik has suggested
\cite{Volovik03} that vortex splitting may have been observed in
\cite{Hoogenboom00}, but he has not presented any calculation to
support this hypothesis.

We model the CuO$_2$ plane as a square array of superconducting
islands. There are two alternative interpretations of this lattice.
Phenomenologically, one may think of it as a coarse-grained model of
the CuO$_2$ plane. Microscopically, adopting the short-range RVB
picture for the sake of simplicity, the lattice may be thought of as
the set of the centers of mass of the nearest-neighbor Cu-Cu singlets.
The fluctuations of the superconducting amplitudes of the islands are
neglected and the only dynamical variable describing the island at $i$
is supposed to be the phase of the condensate $\theta_i$.  The islands
are assumed to be coupled by the Josephson effect and we postulate
that the Hamiltonian of the plane reads as
\begin{equation}
H=\sum_{\langle ij\rangle}
\left[\varepsilon\cos(\theta_i-\theta_j)
-J\cos(2\theta_i-2\theta_j)\right],
\label{eq:josephson_energy}
\end{equation}
where the sum is taken over the nearest-neighbor sites.  The $d$-wave
pairing state is described by $\varepsilon>0$ and in the spin language
it corresponds to an antiferromagnetic configuration of
$\theta_i$. Note that for $\varepsilon<0$ it is the $s$-wave pairing
state that is stable and therefore $\varepsilon=0$ corresponds to a
quantum critical point.  We assume that $J>0$ and therefore the ground
state phase difference in Eq.~(\ref{eq:josephson_energy}) jumps
discontinously from $\pi$ at $\varepsilon>0$ to $0$ at
$\varepsilon<0$.  The alternative choice $J<0$ would correspond to a
continuous change of the ground-state phase difference, thus
physically corresponding to a $d+is$ state in the vicinity of
$\varepsilon=0$, i.e. to a time reversal-breaking mixture of the
$d$-wave and $s$-wave pairing states. Such cooperation of different
pairing states is generically favourable at weak coupling
\cite{Kotliar88}. On the other hand, our case $J>0$ corresponds to a
competition between the pairing states.

In order to proceed we modify the model
Eq.~(\ref{eq:josephson_energy}) in several ways.  First, we perform a
gauge transformation $\theta_i\rightarrow \theta_i+\pi$ on one of the
sublattices. This changes the sign of the first term in
Eq.~(\ref{eq:josephson_energy}) and, as a result, the $d$-wave state
corresponds to the ferromagnetic state. Second, we redefine the zero
of energy so that the homogeneous case corresponds to $E=0$ and
finally, we include the coupling to the magnetic field.  For the sake
of simplicity, we consider a layered tetragonal material with in-plane
lattice constant $d$ and $c$-axis lattice constant $d_c$ and we
consider only vortices along the $c$ axis. Finally we have
\begin{eqnarray}
\tilde{E}=\sum_{\langle ij\rangle}e(\theta_{ij})
+\frac{1}{2}\frac{\lambda^2}{d^2}\sum_i\varphi_i^2,
\label{eq:model_energy}
\end{eqnarray}
where $\tilde{E}=E/(4J+\varepsilon)$ is the dimensionless vortex
energy per CuO$_2$ plane and
$\lambda^{-2}=4\pi^2\mu_0(4J+\varepsilon)/(\Phi_0^2 d_c)$.

The first term in Eq.~(\ref{eq:model_energy}) corresponds to the sum
of dimensionless  Josephson energies of single bonds,
\begin{equation}
e(\theta)=(1-2c)(1-\cos\theta)+\frac{c}{2}(1-\cos 2\theta),
\label{eq:josephson}
\end{equation}
where $c=2J/(4J+\varepsilon)$ is a parameter measuring the strength of
the second harmonic contribution to the Josephson energy of the bonds.
We have introduced a dimensionless vector potential
$a_{ij}=2\pi\int_i^j{\bf A}\cdot d{\bf r}/\Phi_0$ and a gauge
invariant phase difference between lattice sites $i$ and $j$,
$\theta_{ij}=\theta_j-\theta_i+a_{ij}$.  

The second term in Eq.~(\ref{eq:model_energy}) corresponds to the
energy of the magnetic field.  $\varphi_i$ is the dimensionless flux
threading the plaquette with lower left point at $i$. If the plaquette
is formed by the points $ijkl$, then
$\varphi_i=a_{ij}+a_{jk}+a_{kl}+a_{li}$.

Minimizing the energy Eq.~(\ref{eq:model_energy}) with respect to
$a_{ij}$ we obtain the coupled set of discretized Maxwell equations
\begin{eqnarray}
\varphi_{i-\hat{y}}-\varphi_i&=&\frac{d^2}{\lambda^2}
j(\theta_{i+\hat{x}}-\theta_i+a_{i,i+\hat{x}}),
\label{eq:maxwell_x}
\\
\varphi_{i}-\varphi_{i-\hat{x}}&=&\frac{d^2}{\lambda^2}
j(\theta_{i+\hat{y}}-\theta_i+a_{i,i+\hat{y}}),
\label{eq:maxwell_y}
\end{eqnarray}
where $\hat{x}$ and $\hat{y}$ are elementary lattice vectors in the
$x$ and $y$ directions, respectively, and we have introduced the
dimensionless current $j(\theta)=(1-2c)\sin\theta +c\sin 2\theta$.
Note that for slowly varying fields, $|\theta|\ll 1$,
$j(\theta)=\theta$ for all values of $c$. Making use of
Eqs.~(\ref{eq:maxwell_x},\ref{eq:maxwell_y}), we thus identify
$\lambda$ as the penetration depth and from $\lambda\approx
2600\:{\rm\AA}$ \cite{Jacobs95} and $d_c \approx 7.5\:{\rm\AA}$, we
estimate $4J+\varepsilon\approx 6$~meV independently of $c$. Thus $c$
is the only free parameter in the theory.

In what follows we make use of the inequality $\lambda\gg d$ and solve
Eqs.~(\ref{eq:maxwell_x},\ref{eq:maxwell_y}) perturbatively with the
small parameter $d/\lambda$.  To this end let us require that the
phase field $\theta_i$ satisfies the following equations for all sites
$i$:
\begin{equation}
\sum_\tau j(\theta_{i+\tau}-\theta_i)=0,
\label{eq:kirchhoff}
\end{equation}
where the sum is taken over the four nearest neighbor directions
$\tau$.  Physically this corresponds to a lattice version of the
continuity equation $\nabla\cdot{\bf j}=0$ and more formally it might
be thought of as a discretized version of the equation $\nabla^2
\theta=0$. Once Eq.~(\ref{eq:kirchhoff}) is satisfied, one can
estimate the magnetic fields by neglecting the vector potential
$a_{ij}$ on the right-hand sides of
Eqs.~(\ref{eq:maxwell_x},\ref{eq:maxwell_y}) and one finds that
$\varphi_i\propto d^2/\lambda^2\ll 1$.  The vector potential at
distance $R$ from the vortex center can therefore be chosen as $a\sim
(d/\lambda)^2(R/d)$. On the other hand, the typical phase difference
$\theta_{i+\tau}-\theta_i$ at distance $R$ is $d/R$.  Therefore the
current distribution is well described by the phase-only solution
inside the circle with radius $R\sim \lambda$. Beyond this range the
correct solution should differ only marginally from the standard
vortex solution \cite{note}. In this paper we therefore concentrate
only on the region $R\ll \lambda$.

Our task is therefore to find a solution to Eq.~(\ref{eq:kirchhoff})
with a finite winding number. Consider first the standard solution
describing a phase winding by $2\pi$ around the vortex center at
$(0.5,0.5)$.  Consider further the straight line passing through the
vortex center and parallel to the $x$ axis. This line cuts the set of
bonds connecting points $(x,0)$ and $(x,1)$ of the lattice.  We will
work in a gauge where the phase difference
$\Delta\theta_x=\theta_{x0}-\theta_{x1}$ changes between $2\pi$ for
$x\rightarrow -\infty$ and 0 for $x\rightarrow +\infty$.

\begin{figure}[t]
\centerline{\includegraphics[width=0.35\textwidth]{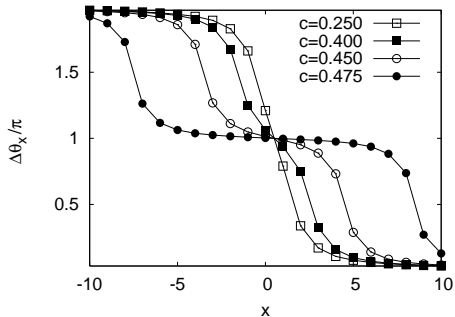}}
\caption{\label{fig:phasejump} Phase difference
$\Delta\theta_x=\theta_{x0}-\theta_{x1}$ for bonds which are cut by a
straight line passing through the vortex center and parallel to the $x$
axis, plotted versus the position $x$ of the bond. The partons are
located at $x=0.5\pm a$. Note the growth of the plateau at
$\Delta\theta_x=\pi$ as the quantum critical point $c=0.5$ is
approached.}
\end{figure}

We have solved Eq.~(\ref{eq:kirchhoff}) on lattices 200$\times$200
numerically by the standard iterative procedure. By varying the
initial configuration, several solutions could be found.  The number
of different solutions increased with $c$. For all solutions we
calculated their Josephson energy and for every studied $c$, we have
identified the optimal solution with minimal energy. From now on, we
will focus on the optimal solutions.

\begin{figure}[h]
\centerline{\includegraphics[width=0.35\textwidth]{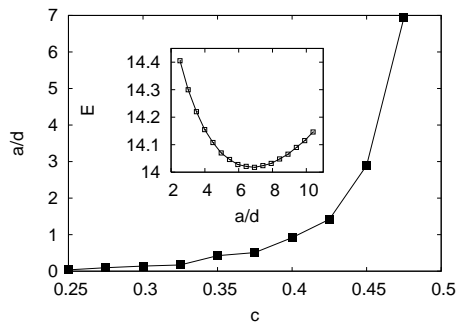}}
\caption{\label{fig:a_c} The equilibrium half-distance $a$ between the
partons (in units of $d$) as a function of $c$. The inset shows the
energy of all solutions found for $c=0.475$ as a function of their
$a$. On the vertical scale, only energy differences are meaningful,
since the energy contains an additive constant dependent on the
lattice size.}
\end{figure}

The results for the phase jump $\Delta\theta_x$ for several values of
$c$ are shown in Fig.~\ref{fig:phasejump}.  In agreement with our
expectations, as a function of $x$, $\Delta\theta_x$ exhibits only a
single step \cite{step} for sufficiently small $c$.  However, for
$c>0.25$ two partial steps develop. The phase jump $\Delta\theta_x$ at
those steps changes from 0 to $\pi$ and from $\pi$ to $2\pi$,
respectively, corresponding to the presence of two partons.  The
spacing between the partons, $2a$, defined as the distance between the
steps, is plotted in Fig.~\ref{fig:a_c}. Note the steep increase of
$a(c)$ for $c$ approaching the quantum critical point at $c=0.5$.

\begin{figure}[h]
\centerline{\includegraphics[width=0.4\textwidth]{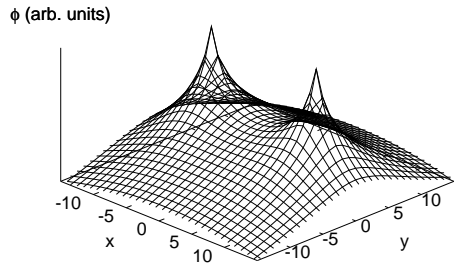}}
\caption{\label{fig:graph1} Spatial distribution of the dimensionless
magnetic flux $\varphi_i$ in a split vortex for $c=0.475$.}
\end{figure}

Once the phase fields are known, the magnetic field distribution can
be calculated making use of
Eqs.~(\ref{eq:maxwell_x},\ref{eq:maxwell_y}). The result of this
calculation is shown in Fig.~\ref{fig:graph1} which explicitly shows
that the vortex splits into two partons.

\begin{figure}[h]
\centerline{\includegraphics[width=0.7\textwidth]{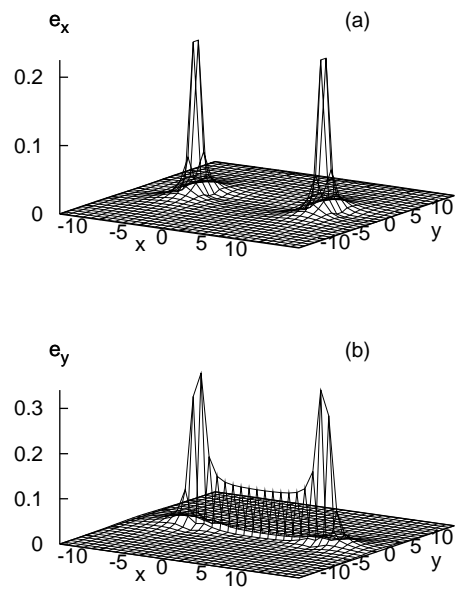}}
\caption{\label{fig:graph2} Kinetic energy (in units of
$4J+\varepsilon$) of the bonds along the $x$ direction (a) and along
the $y$ direction (b) for $c=0.475$.}
\end{figure}

Let us discuss the energetics of vortex splitting in more detail.
Figure~\ref{fig:graph2}a shows the Josephson (kinetic) energy of the
bonds in the $x$ direction. By the standard argument about instability
of multiply charged vortices one can see that this configuration has a
lower total kinetic energy in the $x$ direction, ${\tilde E}_x$, than
the usual vortex solution. Thus, the gradient of ${\tilde E}_x$ pushes
the partons apart \cite{Tinkham96}.

There is however an opposing force, which is caused by the necessary
existence of a cut joining the partons, across which the phase has to
jump approximately by $\pi$.  The kinetic energy of the bonds in the
$y$ direction is shown in Fig.~\ref{fig:graph2}b.  One can see
explicitly that, at long distances $2a$ between the partons, the total
kinetic energy in the $y$ direction, ${\tilde E}_y$, grows linearly
with $a$ due to the energy cost of the cut, thus forming a confining
potential for the partons.

The competition between the repulsive and attractive forces leads to
the presence of a minimum of the total vortex energy as a function of
the parton distance, as demonstrated explicitly in the inset to
Fig.~\ref{fig:a_c}.

So why don't the partons always form? The reason for this is that, for
$c<0.25$, the Josephson energy Eq.~(\ref{eq:josephson}) has a local
maximum at $\theta=\pi$. This destabilizes the parton solution at
those values of $c$. For $c>0.25$, the parton is locally stable, but
the equilibrium interparton distance is in general very short, see
Fig.~\ref{fig:a_c}, because the string tension associated with the cut
is huge. However, when approaching the quantum critical point $c=0.5$,
the energy difference $e(\pi)-e(0)$ measuring the string tension
diminishes and the interparton distance $2a$ grows.

The theory as developed so far applies only to situations when $a\ll
\lambda$.  Therefore it does not apply in the immediate vicinity of
the quantum critical point.  In order to check the robustness of our
picture we have solved the full
equations~(\ref{eq:maxwell_x},\ref{eq:maxwell_y}) at the critical
point $c=0.5$ for a single fractional vortex. To this end we have
rewritten Eqs.~(\ref{eq:maxwell_x},\ref{eq:maxwell_y}) as
\begin{equation}
\varphi_i=\frac{1}{2}\sum_\tau \arcsin
\left[\frac{2\lambda^2}{d^2}(\varphi_{i+\tau}-\varphi_i)
\right] +\pi \delta_{i0},
\label{eq:single}
\end{equation}
where the first term can be thought of as a discretized Laplacian and
only the nearest-neighbor sites contribute to the sum over $\tau$.
The second term is the source term for a vortex whose phase winds by
$\pi$.  The solution to Eq.~(\ref{eq:single}) was obtained by the
standard iterative procedure.  We have checked explicitly that in the
core region $R\ll\lambda$ the approximate solution which satisfies
Eq.~(\ref{eq:kirchhoff}) is in perfect agreement with the full
solution.

The conventional vortex energy at $c=0$ can be estimated using the
London theory \cite{Tinkham96} and we find ${\tilde E}(0)\approx
\pi\ln(\lambda/d)$.  Taking $d$ in the interval between $3.8\:{\rm
\AA}/\sqrt{2}$ and $\xi\approx 14\:{\rm\AA}$ \cite{Ri94}, we estimate
${\tilde E}(0)\approx 16-22$.  From our numerical solution we can
determine the change of the vortex energy with $c$ and, {\it e.g.} at
$c=0.475$, ${\tilde E}(0)-{\tilde E}(0.475)\approx 4.85$. It is seen
that finite values of $c$ may lead to a substantial reduction of the
vortex energy.

Before concluding let us discuss the relevance of our results to the
experiment \cite{Hoogenboom00}. Following \cite{Hoogenboom00}, we
assume that the pinning effects are decisive in determining the vortex
shape.  The difference with respect to \cite{Hoogenboom00} is that in
our picture all vortices in a perfect sample should be split.  The
observation of unsplit vortices can be explained by the attraction of
the partons to the same pinning center by a potential which is
stronger than the energy gain due to splitting \cite{potential}.
Moreover, the interpretation of the observed slow temporal evolution
of the vortex shapes proposed in \cite{Hoogenboom00} is applicable
also in our picture.

In \cite{Hoogenboom00}, the quantum tunneling of vortices between
nearby pinning sites was considered as the most likely mechanism of
vortex delocalization.  Thus the vortex should be described by a
linear superposition of wavefunctions describing the vortex localized
at the various pinning sites.  It is a subtle issue to distinguish a
vortex described by such a wavefunction from our stable parton
picture.  We believe the best way to distinguish these two
alternatives is to determine whether split vortices form also in
perfect samples.

In conclusion, within a simple model we have shown that, in the
vicinity of a quantum phase transition between pairing states of
different symmetry, the vortex cores may acquire a nontrivial parton
structure. We believe that this result is interesting in several
respects: (i) in scanning tunneling spectroscopy of the cuprates, the
parton structure of the vortex cores has been observed in some of the
samples; (ii) the phenomenon is analogous to the formation of the
partial Shockley dislocations; (iii) the parton structure lowers the
energy of the vortices and this may be relevant for the interpretation
of the pseudogap phase; (iv) precisely at the quantum critical point
$\varepsilon=0$, fractional vortices become deconfined.

The author thanks S. Saxena for an interesting discussion.  This work
was supported by the grants APVV-51-003505, VEGA~1/2011/05, and
COST~P-16.



\begin{thebibliography}{99}
\bibitem{step}The step is defined as the locus of steepest descent of
$\Delta\theta_x$ as a function of $x$.
\bibitem{potential}The inset to Fig.~\ref{fig:a_c} shows that the
parton-parton potential is quite soft.
\end{thebibliography}
\end{document}